\documentclass[fleqn]{article}

\def\noi{\noindent}
\newcommand{\eqsection}{\makeatletter
    \@addtoreset{equation}{section}
    \renewcommand{\theequation}{\arabic{section}.\arabic{equation}}
    \makeatother}

\def\lal{&&\nqq {}}

\def\beq{\begin{equation}}
\def\eeq{\end{equation}}
\def\bear{\begin{eqnarray}}
\def\bearr{\begin{eqnarray} \lal}
\def\ear{\end{eqnarray}}
\def\earn{\nonumber \end{eqnarray}}

\def\yy{\\[5pt] {}}

\newcommand{\foom}[1]{\protect\footnotemark[#1]}

\newcommand{\Title}[1]{\noi {\uppercase{\Large #1}} \\}

\newcommand{\Author}[2]{\noi{\large\bf #1}\\[2ex]\noindent{\it #2}\\}

\newcommand{\Acknow}[1]{\subsection*{Acknowledgement} #1}

\newcommand{\Abstract}[1]{\vskip 2mm \begin{center}
        \parbox{16.4cm}{\small\noi #1} \end{center}\medskip}

\newcommand{\email}[2]{\footnotetext[#1]{e-mail: #2}
        \addtocounter{footnote}{1}}

\def\d{\partial}

\def\Jl#1#2{{\it #1\/} {\bf #2},\ }

\def\ApJ#1 {\Jl{Astroph. J.}{#1}}
\def\CQG#1 {\Jl{Class. Quantum Grav.}{#1}}
\def\DAN#1 {\Jl{Dokl. AN SSSR}{#1}}
\def\GC#1 {\Jl{Grav. \& Cosmol.}{#1}}
\def\GRG#1 {\Jl{Gen. Rel. Grav.}{#1}}
\def\JETF#1 {\Jl{Zh. Eksp. Teor. Fiz.}{#1}}
\def\JETP#1 {\Jl{Sov. Phys. JETP}{#1}}
\def\JHEP#1 {\Jl{JHEP}{#1}}
\def\JMP#1 {\Jl{J. Math. Phys.}{#1}}
\def\NPB#1 {\Jl{Nucl. Phys.}{B\ #1}}
\def\NP#1 {\Jl{Nucl. Phys.}{#1}}
\def\PLA#1 {\Jl{Phys. Lett.}{#1A}}
\def\PLB#1 {\Jl{Phys. Lett.}{#1B}}
\def\PRD#1 {\Jl{Phys. Rev.}{D\ #1}}
\def\PRL#1 {\Jl{Phys. Rev. Lett.}{#1}}

\begin{document}

\onecolumn

\Title{CURVATURE COUPLING IN EINSTEIN-YANG-MILLS THEORY \yy AND
NON-MINIMAL SELF-DUALITY}

\Author{A.B. Balakin\foom 1 and A.E. Zayats\foom 2}
{Department  of General Relativity and Gravitation,\\
Kazan State University, 18 Kremlevskaya St, Kazan 420008, Russia}

\Abstract {A self-consistent non-minimal non-Abelian
Einstein-Yang-Mills model, containing three phenomenological
coupling constants, is formulated. The ansatz of a vanishing
Yang-Mills induction is considered as a particular case of the
self-duality requirement for the gauge field. Such an ansatz is
shown to allow obtaining an exact solution of the self-consistent
set of equations when the space-time has a constant curvature. An
example describing a pure magnetic gauge field in the de Sitter
cosmological model is discussed in detail.}


\email 1 {Alexander.Balakin@ksu.ru}%
\email 2 {Alexei.Zayats@ksu.ru}%

\section{Introduction}

The theory of non-minimal coupling of gravity with fields and
media has numerous applications to cosmo\-logy and astrophysics.
Non-minimal theory has been elabora\-ted in detail for scalar and
electromagnetic fields (see, e.g., the review \cite{FaraR} and
references therein). A detailed theory of non-minimal coupling of
gravity with gauge fields is still at its development stage. A
version of the non-minimal Einstein-Yang-Mills (EYM) model was
obtained by M\"uller-Hoissen in 1988 \cite{MH} from dimen\-sional
reduction of the Gauss-Bonnet action. This model contains one
coupling parameter. We follow an alterna\-tive derivation of the
non-minimal EYM theory, formula\-ted as a non-Abelian
generalization of non-minimal non-linear Einstein-Maxwell theory
(see \cite{BL05}) along the line proposed by Drummond and Hathrell
for linear electrodynamics \cite{Drum}. As a particular case, this
theory gives the non-minimal, linear in curvature, EYM model which
can be characterized as a three-parameter model since it contains
three coupling constants $q_1$, $q_2$ and $q_3$. The problem of a
curvature induced back\-reac\-tion of the Yang-Mills field on the
gravitational field seems to be important at least in two aspects.
First, non-minimal coupling of the Yang-Mills field with gravity
can modify the rate of the Universe evolution, providing the
accelerated expansion analogously to the one in the non-minimal
Einstein-Maxwell theory \cite{BZ05}. Second, a curvature coupling
of gauge fields with gravity gives a new degree of freedom in
modeling (regular) spherically symmetric objects \cite{VG}.

In this note, we introduce a three-parameter self-consistent EYM
model in which the EYM Lagrangian satisfies three special
requirements: it is gauge-invariant, linear in space-time
curvature, and quadratic in the Yang-Mills field strength tensor
${\bf F}_{ik}$. Then we consider an exact solution of the obtained
model for a specific case when the non-Abelian induction tensor
${\bf H}_{ik}$ is proportional to the dual field stress tensor
${\bf F}^*_{ik}$, i.e., when a generalized self-duality condition
is satisfied. In this context, we consider an exact solution of
the EYM model with vanishing induction of the gauge field when the
space-time is characterized by a constant curvature and describe
in detail the example of pure magnetic gauge field.

\section{ Non-minimal Einstein-Yang-Mills field equations}

The three parameter non-minimal Einstein-Yang-Mills theory can be
formulated in terms of the action func\-tional
\beq S_{{\rm NMEYM}} = \int d^4 x \sqrt{-g}\ {\cal L}\,,%
\qquad {\cal L}=\frac{R + 2 \Lambda}{\kappa}+
\frac{1}{2}\left[\frac{1}{2}g^{ikmn} {+} {\cal R}^{ikmn}
\right]F^{(a)}_{ik} F_{mn(a)}.
\label{act} \eeq%
Here $\Lambda$ is the cosmological constant, $g = {\rm
det}(g_{ik})$ is the determinant of the metric tensor $g_{ik}$,
$R$ is the Ricci scalar, the constant $\kappa$ is equal to $8\pi
\gamma$, where $\gamma$ is the gravitational constant, Latin
indices without parentheses run from 0 to 3. The symbol $g^{ikmn}$
is a standard abbreviation for the tensor quad\-ratic in the
metric
\begin{equation}
g^{ikmn} \equiv  g^{im}g^{kn} - g^{in}g^{km} \,, \label{1sus}
\end{equation}
the tensor ${\cal R}^{ikmn}$ is defined as follows:
\beq%
{\cal R}^{ikmn} \equiv  \frac{1}{2}\,q_1 R g^{ikmn} +
\frac{1}{2}\,q_2 (R^{im}g^{kn} - R^{in}g^{km} + R^{kn}g^{im}
-R^{km}g^{in}) + q_3 R^{ikmn} \,,
\label{sus} \eeq%
where $R^{ik}$ and $R^{ikmn}$ are the Ricci and Riemann tensors,
respectively, and $q_1$, $q_2$, $q_3$ are the phenomeno\-logi\-cal
parameters describing the non-minimal coupling of the Yang-Mills
and gravitational fields. Following \cite{Rubakov}, we consider
the Yang-Mills field ${\bf F}_{mn}$ taking values in the Lie
algebra of the gauge group $SU(n)$:
\begin{equation}
{\bf F}_{mn} = - i {\cal G} {\bf t}_{(a)} F^{(a)}_{mn} \,, \quad
{\bf A}_m = - i {\cal G} {\bf t}_{(a)} A^{(a)}_m  \,.
\label{represent}
\end{equation}
Here ${\bf t}_{(a)}$ are the Hermitian traceless generators of the
$SU(n)$ group, $A^{(a)}_i$ and $F^{(a)}_{mn}$ are the Yang-Mills
field potential and strength, respectively, the group index $(a)$
runs from $1$ to $n^2-1$, and the constant ${\cal G}$ is the
strength of gauge coupling. Scalar products of the Yang-Mills
fields, indicated by the bold letters, are defined in terms of the
traces of the corresponding matrices (see \cite{Rubakov}), a
scalar product of the generators ${\bf t}_{(a)}$ and ${\bf
t}_{(b)}$ is chosen to be equal to
\begin{equation}
\left( {\bf t}_{(a)} , {\bf t}_{(b)} \right) \equiv 2 {\rm Tr} \
{\bf t}_{(a)} {\bf t}_{(b)}  \equiv G_{(a)(b)}  \,.
\label{scalarproduct}
\end{equation}
The symmetric tensor $G_{(a)(b)}$ plays the role of a metric in
the group space, and the generators can be chosen so that the
metric is equal to the Kronecker delta. The Yang-Mills fields
$F^{(a)}_{mn}$ are connected with the potentials of the gauge
field $A^{(a)}_i$ by the well-known formulae (see, e.g.,
\cite{Rubakov})
\begin{equation}
F^{(a)}_{mn} = \nabla_m A^{(a)}_n {-} \nabla_n A^{(a)}_m + {\cal G}
f^{(a)}_{\cdot (b)(c)} A^{(b)}_m A^{(c)}_n \,. \label{Fmn}
\end{equation}
Here $\nabla _m$ is a covariant space-time derivative, the
sym\-bols $f^{(a)}_{\cdot (b)(c)}$ denote real structure
con\-stants of the gauge group $SU(n)$. The definition of the
commutator in (\ref{Fmn}) is based on the relation
\begin{equation}
\left[ {\bf t}_{(a)} , {\bf t}_{(b)} \right] = i  f^{(c)}_{\cdot
(a)(b)} {\bf t}_{(c)} \,, \label{fabc}
\end{equation}
providing the formula
\begin{equation}
f_{(c)(a)(b)} \equiv G_{(c)(d)} f^{(d)}_{\cdot (a)(b)} = - 2 i \
{\rm Tr} \left[ {\bf t}_{(a)} , {\bf t}_{(b)} \right] {\bf
t}_{(c)}. \label{fabc1}
\end{equation}
The structure constants $f_{(a)(b)(c)}$ are supposed to be
completely antisymmetric under exchange of any two indices
\cite{Rubakov}. The metric $G_{(a)(b)}$ and the structure
con\-stants $f^{(d)}_{\cdot (a)(c) \cdot}$ are supposed to be
constant tensors in the standard and covariant manner. This means
that
\beq%
\d_m G_{(a)(b)} = 0 \,, \quad \hat{D}_m G_{(a)(b)} = 0 \,, \quad
\d_m f^{(a)}_{\cdot (b)(c)} = 0 \,, \quad \hat{D}_m f^{(a)}_{\cdot
(b)(c)} = 0 \,, \label{DfG}
\eeq%
where the following rule for the derivative of the tensors in the
group space is used:
\begin{eqnarray}
\hat{D}_m Q^{(a) \cdot \cdot \cdot}_{\cdot \cdot \cdot (d)} \equiv
\nabla_m Q^{(a \cdot \cdot \cdot)}_{\cdot \cdot \cdot (d)} + {\cal
G} f^{(a)}_{\cdot (b)(c)} A^{(b)}_m Q^{(c) \cdot \cdot
\cdot}_{\cdot \cdot \cdot (d)} - {\cal G} f^{(c)}_{\cdot (b)(d)}
A^{(b)}_m Q^{(a) \cdot \cdot \cdot}_{\cdot \cdot \cdot (c)} \,.
\label{DQ}
\end{eqnarray}

\section{Non-minimal master equations}

\subsection{Non-minimal extension of Yang-Mills equations}

Variation of the action $S_{({\rm NMEYM})}$ with respect to the
Yang-Mills potential $A^{(a)}_i$ yields
\begin{equation}
\hat{D}_k {\bf H}^{ik}  =  0 \,, \quad {\bf H}^{ik} =
\left[\frac{1}{2} g^{ikmn}  + {\cal R}^{ikmn}  \right] {\bf
F}_{mn} \,. \label{HikR}
\end{equation}
The tensor ${\bf H}^{ik}$ is a non-Abelian analogue of the
in\-duction tensor known in the electrodynamics \cite{Maugin,HO}.
This ana\-logy allows us to consider ${\cal R}^{ikmn}$ as a
susceptibi\-lity tensor \cite{BL05}.

The Bianchi identity for the gauge field strength can be written as
\begin{equation}
\hat{D}_k {\bf F}^{*ik} = 0 \,, \label{Aeq2}
\end{equation}
where the asterisk denotes the dual tensor
\begin{equation}
{\bf F}^{*ik} = \frac{1}{2}\epsilon^{ikls} {\bf F}_{ls} \,.
\label{dual}
\end{equation}
Here $\epsilon^{ikls} = \frac{1}{\sqrt{-g}} E^{ikls}$ is the
Levi-Civita tensor while $E^{ikls}$ is the completely
antisymmetric symbol with $E^{0123} = - E_{0123} = 1$.

\subsection{Master equations for the gravitational field}

Variation of the action $S_{({\rm NMEYM})}$ with respect to the
metric yields
\begin{equation}
R_{ik} - \frac{1}{2} R \ g_{ik} = \Lambda \ g_{ik} + \kappa
T^{({\rm eff})}_{ik} \,. \label{Ein}
\end{equation}
The effective stress-energy tensor $T^{({\rm eff})}_{ik}$  can be
divided into four parts:
\begin{equation}
T^{({\rm eff})}_{ik} =  T^{(YM)}_{ik} + q_1 T^{(I)}_{ik} + q_2
T^{(II)}_{ik} + q_3 T^{(III)}_{ik} \,. \label{Tdecomp}
\end{equation}
The first term
\begin{equation}
T^{(YM)}_{ik} \equiv \frac{1}{4} g_{ik} F^{(a)}_{mn}F^{mn}_{(a)} -
F^{(a)}_{in}F_{k (a)}^{\cdot n} \label{TYM}
\end{equation}
is the stress-energy tensor of the pure Yang-Mills field. The
definitions of the other three tensors are related to the
corresponding coupling constants $q_1$, $q_2$, $q_3$:
\beq%
T^{(I)}_{ik} = R\,T^{(YM)}_{ik} -  \frac{1}{2} R_{ik}
F^{(a)}_{mn}F^{mn}_{(a)} + \frac{1}{2} \left[ \hat{D}_{i}
\hat{D}_{k} - g_{ik} \hat{D}^l \hat{D}_l \right]
\left[F^{(a)}_{mn}F^{mn}_{(a)} \right] \,, \label{TI}
\eeq%

\[%
 T^{(II)}_{ik} = -\frac{1}{2}g_{ik}\biggl[\hat{D}_{m}
\hat{D}_{l}\left(F^{mn (a)}F^{l}_{\ n(a)}\right)-R_{lm}F^{mn (a)}
F^{l}_{ \ n (a)} \biggr] {-} F^{ln}_{(a)} \left(R_{il}F^{(a)}_{kn}
+ R_{kl}F^{(a)}_{in}\right) \]

\beq%
 -R^{mn}F^{(a)}_{im} F_{kn (a)} -
\frac{1}{2} \hat{D}^m \hat{D}_m \left(F^{(a)}_{in} F_{k(a)}^{ \
n}\right)+ \frac{1}{2}\hat{D}_l \left[ \hat{D}_i \left(
F^{(a)}_{kn}F^{ln}_{(a)} \right) + \hat{D}_k
\left(F^{(a)}_{in}F^{ln}_{(a)} \right) \right] \,, \label{TII}
\eeq%

\beq%
T^{(III)}_{ik} = \frac{1}{4}g_{ik} R^{mnls}F^{(a)}_{mn}F_{ls(a)}-
\frac{3}{4} F^{ls}_{(a)} \left(F_{i}^{\cdot n (a)} R_{knls} +
F_{k}^{\cdot n (a)}R_{inls}\right) -\frac{1}{2}\hat{D}_{m}
\hat{D}_{n} \left[ F_{i}^{ \ n (a)}F_{k(a)}^{ \ m} + F_{k}^{ \ n
(a)} F_{i(a)}^{ \ m } \right] \,. \label{TIII}
\eeq%

\subsubsection{Trace of the effective stress-energy ten\-sor}

In contrast to the traceless stress-energy tensor of the
Yang-Mills field $T^{(YM)}_{ik}$, the stress-energy tensors
$T^{(I)}_{ik}$, $T^{(II)}_{ik}$, $T^{(III)}_{ik}$ have
non-vanishing traces:
\beq%
T^{(I)} =  - \frac{1}{2} \left( R + 3 \hat{D}^{l} \hat{D}_{l}
\right) \left[F^{(a)}_{mn}F^{mn}_{(a)}\right] \,, \label{TIspur}
\eeq%

\beq%
T^{(II)} = - \hat{D}^{m} \hat{D}^{k}\left(F_{mn (a)} F^{ \ n
(a)}_{k} \right) - \frac{1}{2} \hat{D}^k \hat{D}_k
\left(F^{(a)}_{mn}F^{mn}_{(a)} \right) - R^{mn}g^{ik} F^{(a)}_{im}
F_{kn(a)} \,, \label{TIIspur}
\eeq%

\beq T^{(III)} = - \frac{1}{2} R^{mnls} F^{(a)}_{mn}F_{ls(a)} -
\hat{D}_{m} \hat{D}_{n} \left[F^{kn (a)}F_{k(a)}^{ \ m }
\right]\,. \label{TIIIspur}\eeq

\subsubsection{Bianchi identities}

The right-hand side of the Einstein equations (\ref{Ein}) must be
divergence-free. This is valid automatically if $F^{(a)}_{ik}$ is
a solution to the Yang-Mills equations (\ref{HikR}) and
(\ref{Aeq2}). To check this fact directly, one has to use the
Bianchi identities and the properties of the Riemann tensor:
\beq%
\nabla_i R_{klmn} + \nabla_l R_{ikmn} + \nabla_k R_{limn} = 0\,,
\quad R_{klmn} + R_{mkln} + R_{lmkn} = 0 \,, \label{bianchi}
\eeq%
as well as the commutation rules for covariant deriva\-tives \beq
(\nabla_l \nabla_k - \nabla_k \nabla_l) W^i = W^m R^i_{\cdot mlk}
\,. \label{nana} \eeq

Note that the susceptibility tensor ${\cal R}_{ikmn}$ has the same
symmetry of indices as the Riemann tensor, more\-over, the second
identity in (\ref{bianchi}) will be valid if the Riemann tensor is
replaced with ${\cal R}_{ikmn}$.

\section{Generalized self-duality and exact solutions in the EYM model}

The well-known self-duality problem for the Yang-Mills field (see,
e.g., \cite{Belavin}) can be generalized to the case when, in
addition to the gauge field strength, one has a gauge field
induction. When
\beq%
H_{ik}^{(a)} = \lambda_{(b)}^{(a)} F^{*\,(b)}_{ik}\,,\label{HLF}
\eeq%
the Yang-Mills equations (\ref{HikR}) are satisfied due to
(\ref{Aeq2}), when $\hat{D}_m\lambda_{(b)}^{(a)}=0$. The latter
requirement is valid for arbit\-rary ${\bf A}_m$ if
\beq%
\d_m\lambda_{(b)}^{(a)}=0\,, \quad
f^{(a)}_{(c)(d)}\lambda_{(b)}^{(d)}=f^{(d)}_{(c)(b)}\lambda_{(d)}^{(a)}\,.\label{Lab}
\eeq%
For the SU(n) gauge group, the conditions (\ref{Lab}) yield
\beq%
\lambda_{(b)}^{(a)}=\lambda\,\delta_{(b)}^{(a)}\,, \quad \lambda
\equiv \frac{1}{(n^2-1)}\lambda_{(a)}^{(a)}\,.
\eeq%
For arbitrary gauge groups, the matrix $\lambda_{(b)}^{(a)}$ is
not necessarily diagonal (see, e.g., \cite{Zayats}).

\par Thus, for SU(n) symmetry, the Eq.(\ref{HLF}) takes the
form
\begin{equation}
\lambda {\bf F}^{*}_{ik} = {\bf F}_{ik} + {\cal R}_{ikmn} {\bf
F}^{mn} \,. \label{main}
\end{equation}
Second dualization of the relation (\ref{main}) gives
\begin{equation}
{\bf F}_{ik}(1{+}\lambda^2) {+} {\cal R}_{ikmn} {\bf F}^{mn} {+}
^{*}{\cal R}_{ikmn} \lambda {\bf F}^{mn} = 0 \,. \label{ma}
\end{equation}
In the minimal EYM theory with ${\cal R}_{ikmn} =0$, the relation
(\ref{ma}) requires that ${\bf F}_{ik}(1{+}\lambda^2)=0$ and thus
${\bf F}_{ik}=0$ if $\lambda$ is a real constant. If ${\cal
R}_{ikmn}\neq 0$, a non-trivial solution for ${\bf F}_{ik}$ can
exist.

The relation (\ref{main}) generalizes the well-known self-duality
condi\-tion in electrodynamics and Yang-Mills theory. We
distinguish two different cases. In the first one, the relation
(\ref{main}) is satisfied for arbitrary ${\bf F}_{ik}$ due to a
special choice of the coupling constants $q_1$, $q_2$, $q_3$ and
the symmetry of space-time. In the second case, the relation
(\ref{main}) is satisfied for a specific structure of ${\bf
F}_{ik}$. In this note we focus on the first case only.

\subsection{Non-minimal EYM models with vanishing induction}

The equation (\ref{main}) is satisfied for arbitrary ${\bf
F}_{ik}$ if
\beq%
\frac{\lambda}{2}\, \epsilon^{ikmn} =
\frac{1}{2}\,g^{ikmn} + {\cal R}^{ikmn}. \label{22}
\eeq%
A cyclic transposition of the last three indices in (\ref{22})
yields $\lambda=0$, providing the relation
\beq%
{\cal R}_{ikmn}= - \frac{1}{2} \, g_{ikmn}. \label{main2}
\eeq%
Direct consequences of (\ref{main2}) are
\beq%
\frac{3q_1+q_2}{2} R g_{kn} + \left(q_2+q_3\right) R_{kn} = -
\frac{3}{2} g_{kn} \,, \label{25}
\eeq%
\beq%
\left(6q_1+3q_2+q_3\right)R=-6 . \label{251}
\eeq%
When $6q_1+3q_2+q_3\neq 0$, one obtains
\beq%
R = - 12 K \,, \quad K \equiv \frac{1}{2(6q_1+3q_2+q_3)}.
\label{252}
\eeq%
If  $q_2+q_3 \neq 0$, the Ricci tensor $R_{kn}$ can be extracted
from (\ref{25}):
\beq%
R_{kn} = - 3Kg_{kn}\,. \label{253}
\eeq%
Similarly, if  $q_3 \neq 0$, the Riemann tensor can be obtained
from (\ref{main2}):
\beq%
R_{ikmn} = - K g_{ikmn} \,. \label{254}
\eeq%
Eqs.(\ref{252})-(\ref{254}) show that a self-dual non-minimal EYM
model with a vanishing induction tensor requires the space-time to
be of constant curvature $K$. The condi\-tion $K=0$ is
incompatible with (\ref{22}). Thus, varying the parameters $q_1$,
$q_2$, $q_3$, one can formulate five different submodels. The
first submodel is the general one with $q_3 \neq0$,
$q_2+q_3\neq0$, $6q_1+3q_2+q_3 \neq 0$. These conditions allow one
to find $R$, $R_{ik}$ and $R_{ikmn}$ unambiguously in the form
(\ref{252})-(\ref{254}). The second (first special) submodel with
$q_3=0$, $q_2\neq0$, $2q_1+q_2\neq 0$, allows one to find the
Ricci tensor in the form (\ref{253}) and Ricci scalar according to
(\ref{252}), but the Riemann tensor itself cannot be extracted.
The third  (the second special) submodel with $q_2+q_3 =0$, $q_3
\neq0$, $3q_1+q_2\neq 0$ gives the Ricci scalar according to
(\ref{252}) and the Riemann tensor in the form
\beq
R_{ikmn}=\frac{1}{2}(R_{im}g_{kn}+R_{kn}g_{im}-R_{in}g_{km}-R_{km}g_{in})+\frac{1}{2\left(3q_1+q_2\right)}
g_{ikmn} \,, \label{255} \eeq the Ricci tensor being unresolvable.
Note that the rela\-tion (\ref{255}) is valid when the Weyl tensor
vanishes. The last special model with $q_2=q_3=0$, $q_1\neq0$
yields $R=-\frac{1}{q_1}$, but $R_{ikmn}$ and  $R_{kn}$ cannot be
identified.

\subsection{Example of an exact solution}

The self-duality discussed above guarantees the non-minimal
Yang-Mills equations to be satisfied identically for an arbitrary
potential ${\bf A}_i$ if the relation (\ref{main2}) is valid. To
solve the whole set of non-minimal EYM field equations, one needs
to consider the remaining equations (\ref{Ein})-(\ref{TIII}). In
this note we consider a de Sitter-type model with $q_1=q_2=0$ and
the SU(2)-symmetric gauge field. This model is related to the
non-vanishing constant cur\-va\-ture $K = \frac{1}{2q_3}$ and is a
non-Abelian generalization of Prasanna's electrodynamic model
\cite{Prasa1}. For this model, the gravitational field equations
reduce to
\bear%
\left(3K - \Lambda \right)g_{ik} = - \frac{\kappa}{2K}\hat{D}^m
\hat{D}^n \left(F_{in}^{(a)} F_{km(a)} \right) +\frac{\kappa}{2}\,
g^{mn}F_{in}^{(a)}F_{km(a)}\,.\label{3}
\ear%
In the static representation, de Sitter space-time is
characterized by the metric
\bear%
ds^2=\left(1-Kr^2 \right) dt^2-\frac{dr^2}{(1-Kr^2)}-r^2[d\theta^2
+ \sin^2\theta d\varphi^2] \,. \label{deS}
\ear%
Consider the so-called pure magnetic solution with the following
ansatz (see, e.g., \cite{VG})
\beq%
{\bf A}_{0} = {\bf A}_{r} = 0\,,\quad {\bf
A}_{\theta}=-i\left(w(r)-1\right) \ {\bf t}_{\varphi} \,, \quad
{\bf A}_{\varphi}=i\left(w(r)-1\right)\sin{\theta} \ {\bf
t}_{\theta} \,. \label{deS2}
\eeq%
Here ${\bf t}_r$, ${\bf t}_{\theta}$ and ${\bf t}_{\varphi}$ are
the position-dependent gene\-rators of the  SU(2) group:
\[%
{\bf t}_r=\cos{\varphi}\sin{\theta}\;{\bf
t}_{(1)}+\sin{\varphi}\sin{\theta}\;{\bf
t}_{(2)}+\cos{\theta}\;{\bf t}_{(3)},
\]%
\beq%
{\bf t}_{\theta}=\partial_{\theta}{\bf t}_r,\qquad {\bf
t}_{\varphi}=\frac {1}{\sin{\theta}}\ \partial_{\varphi}{\bf t}_r,
\label{deS5}
\eeq%
which satisfy the relations
\beq%
\left[{\bf t}_{r},{\bf t}_{\theta}\right]=i\,{\bf
t}_{\varphi},\quad \left[{\bf t}_{\theta},{\bf
t}_{\varphi}\right]=i\,{\bf t}_{r}, \quad \left[{\bf
t}_{\varphi},{\bf t}_{r}\right]=i\,{\bf t}_{\theta}.\label{deS6}
\eeq%
Non-vanishing components of the field strength tensor are
\beq%
{\bf F}_{r \theta}=-iw'\;{\bf t}_{\varphi}\,, \quad {\bf
F}_{r\varphi}=iw'\sin{\theta}\;{\bf t}_{\theta}\,,\quad {\bf
F}_{\theta \varphi}=-i\left(w^2-1\right)\sin{\theta}\;{\bf t}_r
\,. \label{573}
\eeq%
The gravity field equations yield, first, the standard relation
between cosmological constant and curvature, $\Lambda=3K$, and,
second, the only non-trivial differential equa\-tion for the
function $w(r)$
\beq%
\left(\frac{{(1-Kr^2)w'}^2}{r}\right)' = \frac{(w^2-1)^2}{r^4}
\label{574}
\eeq%
(the prime denotes a derivative with respect to $r$). Thus, the
whole set of differential equations in the non-minimal EYM model
is reduced to a single second\-order differen\-tial equation. It
has the constant solu\-tion ${w(r)=\pm 1}$, however, it describes
a pure gauge (${\bf F}_{ik}=0$). Another exact solution expressed
in terms of elementary functions is
\beq%
w(r)=\pm\sqrt{1-Kr^2}\,. \label{ExSol}
\eeq%
This solution is regular at the origin for arbitrary $K$. When $K$
is positive, the metric (\ref{deS}) has a horizon at
$r_h=\frac{1}{\sqrt{K}}$ and $w(r_h)=0$. Other solutions of
Eq.(\ref{574}) can be found numerically.

\section{Discussion}

One of the results presented in the paper is a deriva\-tion of a
new self-consistent non-minimal set of master equations for the
coupled Yang-Mills and gravity fields from a gauge-invariant
non-minimal Lagrangian. The obtained mathemati\-cal model contains
three arbitrary parameters, and thus admits a wide choice of
special submodels inte\-resting for applica\-tions to non-minimal
cosmo\-logy (isotropic and ani\-sotropic) and non-minimal coloured
spheri\-cally symmetric ob\-jects.

\par To present an exact solution of the formula\-ted model,
we have considered the ansatz of self-duality of the Yang-Mills
field and focused attention on its particular case, a model with
vanishing gauge field induction. This ansatz happens to be
compatible with the non-minimal EYM model, when the space-time is
characteri\-zed by a constant curvature. To show that this model
contains at least one non-trivial solution, we have considered an
example describing a pure magnetic non-Abelian gauge field in de
Sitter space-time. We have shown that the whole set of equations
reduces to one differential equation for one required function.
This circumstance guarantees the consistency of the model. A new
exact solution expressed in terms of elementary functions regular
both at the origin and at the horizon is obtained.

\par Since the model contains three arbitrary parameters (coupling
constants), there arises the problem of intro\-duction of ``new
constants of Nature'' with the dimen\-sionality of area. We show
that the coupling constants introduced phenomenologically can be
inter\-preted in terms of the cosmological constant.

\Acknow This work was supported by the Deutsche
Forschungs\-gemeinschaft.


\small
\begin{thebibliography}{99}

\bibitem{FaraR} V. Faraoni, E. Gunzig and P. Nardone, {\it Fundamentals of Cosmic
Physics} {\bf 20}, 121 (1999).

\bibitem{MH} F. M\"uller-Hoissen, \CQG 5 L35 (1988).

\bibitem{BL05} A.B. Balakin and J.P.S. Lemos, \CQG 22 1867 (2005).

\bibitem{Drum} I.T. Drummond and S.J. Hathrell, \PRD 22 343 (1980).

\bibitem{BZ05} A.B. Balakin and W. Zimdahl, \PRD 71 124014 (2005).

\bibitem{VG} M.S. Volkov and D.V. Gal'tsov, {\it Phys. Rept.} {\bf 319}, 1 (1999).

\bibitem{Rubakov} V. Rubakov, ``Classical Theory of Gauge
Fields'', Princeton University Press, Princeton and Oxford, 2002.

\bibitem{Maugin} A.C. Eringen and G.A. Maugin, ``Electrodynamics of
continua'', Springer-Verlag, New York, 1989.

\bibitem{HO} F.W. Hehl and Yu.N. Obukhov, ``Foundations of Classical
Electrodynamics: Charge, Flux, and Metric'', Birkha\"user, Boston,
2003.

\bibitem{Belavin} A.A. Belavin, A.M. Polyakov, A.S. Schwartz and
Yu.S. Tyupkin, \PLB 59 85 (1975).

\bibitem{Zayats} A.E. Zayats, {\it in: \/} ``Recent problems in field theory'',
v.~4, ed. A.V. Aminova, Heter, Kazan, 2004.

\bibitem{Prasa1} A.R. Prasanna, \PLA 37 331 (1971).

\end{thebibliography}
\end{document}